\documentclass[superscriptaddress,floatfix,aps,pre,amsmath,amssymb,twocolumn,shortbibliography]{revtex4-2}

\usepackage[utf8]{inputenc}
\usepackage{siunitx}
\usepackage{graphicx}
\usepackage{hyperref}
\hypersetup{colorlinks,allcolors=black}
\usepackage{tcolorbox}

\usepackage[capitalise]{cleveref}
\usepackage{microtype}
\usepackage{bm}
\usepackage{lipsum}
\usepackage[english]{babel}
\usepackage{color}
\usepackage{graphicx}
\usepackage{xcolor}
\usepackage{braket}

\newcommand{\bp}{\mathbf{p}}

\newcommand{\Lagr}{\mathcal{L}}

\begin{document}

\author{Sebastian Lundström}
\email{sebastian.lundstrom@umu.se}
\affiliation{Department of Physics, Ume{\aa} University, SE--901 87 Ume{\aa}, Sweden}

\author{Philip Semr\'{e}n}
\affiliation{Department of Physics, Ume{\aa} University, SE--901 87 Ume{\aa}, Sweden}

\title{Modified vacuum polarization in the presence of a plasma}

\author{Haidar Al-Naseri}
\affiliation{Stanford PULSE Institute, SLAC National Accelerator Laboratory, Menlo Park, California 94025, USA}
\affiliation{Department of Physics, University of Gothenburg, 41296 Gothenburg, Sweden}

\author{Gert Brodin}
\email{gert.brodin@umu.se}
\affiliation{Department of Physics, Ume{\aa} University, SE--901 87 Ume{\aa}, Sweden}

\begin{abstract}
We study vacuum polarization due to strong fields, in the presence of an electron-positron plasma. For this purpose, we expand quantum kinetic equations using weak fields and slow temporal scales as expansion parameters. It is demonstrated that the evolution of the Dirac field can be described by classical-like distribution functions for electrons and positrons, which are weakly coupled through quantum interactions. Furthermore, we deduce that these coupling terms give rise to well-known expressions for vacuum polarization, in addition to quantum modifications proportional to the content of real particles. Depending on the initial plasma density, the dominant quantum corrections to classical evolution may arise from real particle couplings or from the vacuum polarization associated with virtual particles. The implications of our results are discussed.    
\end{abstract}
 
\maketitle

\section{Introduction}
Recently, there have been many theoretical treatments that generalize classical descriptions of plasmas to cover quantum mechanical effects, see e.g. \cite{vladimirov2011description,brodin2022quantum,manfredi2021fluid,melrose2020quantum}. The need for quantum treatments could be due to high plasma densities and/or low temperatures \cite{vladimirov2011description,brodin2022quantum,manfredi2021fluid,melrose2020quantum}, or it could be due to strong electromagnetic fields \cite{brodin2022quantum,marklund2006nonlinear,fedotov2023advances,gonoskov2022charged,di2012extremely}. Strong field physics could be of interest in astrophysical situations (where, in magnetars, magnetic field strengths may be comparable to the critical field \cite{harding2006physics}), or in laboratory laser-plasma interaction. In the latter context, we note that the recent evolution of high-power lasers has led to a host of different quantum phenomena becoming accessible to experiments \cite{fedotov2023advances,gonoskov2022charged,di2012extremely}. 

Theoretical treatments of quantum plasmas have normally been chosen, depending on whether you are focusing on the high-density low-temperature regime, or the strong field regime. In the former case, you can focus on quantum effects that enter when the characteristic de Broglie length approaches the Debye length. In the latter case, the quantum $\chi$-parameter (electromagnetic field over the Schwinger critical field in the rest frame of the particle) is typically the relevant measure of the magnitude of quantum phenomena. However, while there naturally are theoretical justifications to apply different expansion parameters depending on the system of study, there are phenomena tying strong field and low temperature (and high density) physics together. For example, spin polarization of a plasma may occur due to a strong laser field \cite{del2017spin,del2018electron,li2019ultrarelativistic}, or it may be related to a low temperature, since the degree of spin polarization in a background magnetic field is inversely proportional to the temperature \cite{hussain2014weakly,andreev2015separated,manfredi2019phase}. 

In some cases, there is the possibility of adding strong field phenomena to an otherwise classical plasma description in a relatively simple way. Specifically, even in the presence of a plasma, a strong electromagnetic field will induce vacuum polarization. An expression for a current density associated with strong field vacuum polarization can then be superimposed on particle plasma currents, calculated from classical theory. Accordingly, Ref. \cite{medvedev2023plasma} has used a setup with plasma currents described by the relativistic Vlasov equation, combined with vacuum polarization currents described by the Euler-Heisenberg Lagrangian \cite{marklund2006nonlinear,dittrich2001probing}, which was used to calculate a conductivity tensor in the presence of a strong magnetic field. In a similar spirit, Ref. \cite{huang2015quantum} combined classical plasma currents with vacuum polarization to describe birefringence in a pair plasma, Ref. \cite{lai2003transfer} combined classical vacuum susceptibilities with vacuum polarization contributions to describe astrophysical phenomena,  and Ref. \cite{chen2013nonlinear} studied nonlinear plasma dynamics using the same type of approach. Moreover, to describe strong field plasma phenomena, many other works (see e.g. Refs. \cite{qu2021signature,melrose2013qpd2,heyl1999nonlinear,marklund2006nonlinear,lundin2010qed}) have also added vacuum polarization currents to particle plasma currents, where the latter have been described with or without quantum corrections.

While there are situations where you expect a simple addition of plasma current and vacuum polarization currents to work on physical grounds, a theoretically more satisfactory approach is to derive the vacuum and particle currents from a unified approach. In this way, the validity of the approach would be apparent, and possible quantum corrections to the particle currents would enter in a consistent way. The so-called Dirac-Heisenberg-Wigner (DHW) formalism \cite{bialynicki1991phase,hebenstreit2010schwinger,ilderton2012lightfront}, which gives a phase-space description of the Dirac theory, offers the possibility of treating particle currents quantum relativistically, while at the same time including the effects of vacuum polarization in a unified framework. 

In the present work, we will apply the DHW formalism in the specific case of a restricted field geometry to derive evolution equations that simultaneously include quantum-corrected particle currents as well as vacuum polarization currents described by the Euler-Heisenberg Lagrangian. Here, the vacuum polarization appears as a built-in feature of the DHW formalism, and is not added by hand. The results are analyzed,  and for the specific field geometry considered, it is found that vacuum polarization dominates over quantum corrections involving real particles in the extreme ultra-relativistic regime. Specifically, vacuum polarization currents give a larger contribution for gamma factors fulfilling $\gamma\gtrsim 300$. 

The organization of the paper is as follows: In \cref{Sec: II}, we give a brief review of vacuum polarization. \cref{Sec: III} presents the DHW formalism in the field geometry of consideration, introduces the expansions in weak field strengths and long scale lengths, and describes the charge renormalization. In \cref{Sec: Kinetic theory for electrons and positrons}, we demonstrate that the evolution equation can be described in terms of classical-like distribution functions for electrons and positrons, obeying relativistic Vlasov equations, with coupling terms due to spin polarization and the quantum vacuum. Next, in \cref{Sec: V}, the homogeneous limit is analyzed. Finally, in \cref{Sec: VI}, the consequences of our findings are discussed.
\section{Background}  \label{Sec: II}

The physical system to be studied in the sections below, specifically the Dirac field under the influence of dynamically varying electric fields, will include effects due to the quantum vacuum, while at the same time keeping effects of the same order associated with real electrons and positrons. For later comparisons,  we here give a brief review of current sources in vacuum that are induced by electromagnetic fields interacting with the quantum vacuum, in the absence of real particles. 

In this section, we will not make derivations from first principles, but rather present some important standard results.  The main assumption is that the field strength is much lower than the Schwinger critical field. Keeping terms up to fourth order in the electromagnetic field strength, vacuum polarization can be derived from the following form of the Euler-Heisenberg Lagrangian density \cite{marklund2006nonlinear,dittrich2001probing,eriksson2004possibility}
\begin{equation}
\Lagr _{EH}=\mathcal{F}+\frac{\alpha }{90\pi E_{cr}^{2}}\left( 4\mathcal{F}^{2}%
\mathcal{+}7\mathcal{G}^{2}\right), 
\end{equation}%
where the electromagnetic invariants are 
\begin{equation}
\mathcal{F}=\frac{1}{4}F_{\mu
\nu }F^{\mu \nu }=\frac{1}{2}(E^{2}-B^{2})
\end{equation}
and 
\begin{equation}
\mathcal{G=(}1/4\mathcal{)}F_{\mu
\nu }\hat{F}^{\mu \nu }=\mathbf{E}\mathbf{\cdot B}
\end{equation} 
where $\hat{F}^{\mu \nu
}$ is the dual of $F^{\mu \nu }.$ 
Moreover, $\alpha$ is the fine-structure constant, $E_{cr}=m^2c^3/e\hbar$ is the Schwinger critical field, where $m$ is the electron mass, $e$ is the elementary charge, $c$ is the speed of light in vacuum, and $\hbar$ is the reduced Planck's constant. 

Varying the four-potential, minimizing the
action, only including the first term of the Lagrangian density ($%
\propto \mathcal{F}$), leads to the standard (classical) form of Maxwell's
equations in vacuum. Including the terms $\propto \mathcal{F}^{2}$ and $%
\mathcal{G}^{2}$ resulting from the interaction with the quantum vacuum,
adds the effect of vacuum polarization. That is, we get vacuum source terms
in Maxwell's equations with current densities ${\bf j}$ and charge densities $\rho$ given by the
standard forms 
\begin{eqnarray}
\mathbf{j} &=&\frac{\partial \mathbf{P}}{\partial t}+\nabla \times \mathbf{M},
\\
\rho  &=&-\nabla \cdot \mathbf{P},
\end{eqnarray}
and where the vacuum polarization $\mathbf{P}$ and vacuum magnetization $%
\mathbf{M}$ are given by%
\begin{eqnarray}
\mathbf{P} &\mathbf{=}&\frac{\alpha }{45\pi E_{cr}^{2}}\left[ 2(E^{2}-B^{2})%
\mathbf{E}+7\mathbf{E}\mathbf{\cdot BB}\right]   \label{polarization} ,\\
\mathbf{M} &\mathbf{=}&\frac{\alpha }{45\pi E_{cr}^{2}}\left[ \mathbf{-}%
2(E^{2}-B^{2})\mathbf{B}+7\mathbf{E}\mathbf{\cdot BE}\right].
\label{magnetization}
\end{eqnarray}%
\cref{polarization,magnetization} can be used to derive, for example, cross-sections for
photon-photon interaction in vacuum, expressions for photon splitting in the
presence of a magnetic field, self-focusing of electromagnetic radiation,
and many other phenomena, see e.g. Refs. \cite{marklund2006nonlinear,dittrich2001probing}. 

Although $\Lagr_{EH}$ in the form given here has been derived under the
condition of fields weaker than the critical field, the extra terms in the
Lagrangian density are of higher order in the field strength, and thus the above
expressions for $\mathbf{P}$ and $\mathbf{M}$ are nevertheless strong-field
effects.  In addition to these strong field contributions, there is a
correction term to the Lagrangian density associated with rapidly varying fields
(see e.g. Refs. \cite{marklund2006nonlinear,mamaev1981effective}). Note that although we speak of "rapidly varying fields", we still refer to scale lengths that are much longer than the Compton scale. If we include correction terms in the
Lagrangian density, associated with rapidly varying fields, we get the following expression   
\begin{equation}
\Lagr=\Lagr_{EH}+\Lagr_{D},
\end{equation}%
where the contribution due to rapid variations is%
\begin{equation}
\Lagr_{D}=\frac{2\alpha }{15\omega _{c}^{2}}\left[ \partial _{a}F^{ab}\partial
_{c}F_{b}^{c}-F_{ab}\square F^{ab}\right], 
\end{equation}%
where $\square =\partial _{a}\partial ^{a}$ is the d'Alembertian \cite{mamaev1981effective,marklund2006nonlinear} and $\omega_c =mc^2/\hbar$ is the Compton frequency. Only
considering the short scale corrections (dropping terms $\propto \mathcal{F}%
^{2}$, $\mathcal{G}^{2}$), the four-current density $j^b$ added to the previously
given vacuum polarization source is given by  
\begin{equation}
j^{b}=\frac{2\alpha }{15\omega _{c}^{2}}\partial _{a}F^{ab}.
\label{derivative-corr}
\end{equation}%
Effects due to Eq.~(\ref{derivative-corr}) has been studied in, for example,
Refs \cite{marklund2006nonlinear,lundin2007short, rozanov1998self}. A key property of the short-scale correction given by Eq.~(\ref{derivative-corr}) is that the contribution adds
dispersion to wave modes that otherwise would be dispersion-free. While the
current sources in \cref{polarization,magnetization} and \cref{derivative-corr} originally have been derived for the case of vacuum, naturally these
effects are still present in a plasma of sufficiently low density. 
Thus,
many works (e.g. \cite{medvedev2023plasma,lai2003transfer,chen2013nonlinear,huang2015quantum,qu2021signature,melrose2013qpd2,heyl1999nonlinear,marklund2006nonlinear,lundin2010qed}) have studied systems where vacuum contributions are added to
plasma currents. While this is feasible in many
cases, as we will see below, depending on the plasma density, a simple superposition of vacuum
currents and classical currents associated with real particles may not always work. 

\section{Basic equations and expansion parameters} \label{Sec: III}

In this section, we will introduce the DHW formalism, see e.g. \cite{bialynicki1991phase,hebenstreit2010schwinger,ilderton2012lightfront}, for a special case. Then, we will expand the DHW equations in weak fields (compared to the critical field) and slow time-scales (compared to the Compton scale). This expansion will be used to derive an evolution equation for the electric field in the homogeneous limit, and to obtain coupled equations for classical-like distribution functions of electrons and positrons in the weakly inhomogeneous limit. Due to the restrictions imposed in the field geometry, the field magnitude, and in the temporal and spatial scales, the dynamics will be classical to leading order. However, we note that the expansion in the field magnitude and temporal and spatial scales alone does not necessarily imply classical dynamics. The purpose of nevertheless restricting the field geometry is to allow for a straightforward and mostly analytical comparison of quantum corrections from vacuum polarization and quantum corrections due to the presence of matter.    

\subsection{The DHW formalism in the electrostatic limit}
In this subsection, we will introduce the DHW formalism applied to a case of reduced field geometry. Our starting point is the DHW formalism, firstly developed in Ref. \cite{bialynicki1991phase}. The general governing equations are derived by applying the Wigner operator based on the Dirac 4-spinors, using the Wilson line factor to ensure gauge invariance (as first introduced in Ref.  \cite{Stratonovich1956Dokl}), and making a Hartree (mean field) approximation. The resulting 16 scalar equations are then split into quantities representing phase-space functions for spin-density, current density, mass density, etc. Good accounts of the general formalism are given in the original source \cite{bialynicki1991phase} as well as some other works, e.g. Refs. \cite{hebenstreit2010schwinger,sheng2019wigner}.  

Here, we will restrict ourselves to a case with restricted field geometry, such that the magnetic field ${\bf B}\equiv 0$, and that the electric field $\bf E$ depends on a single spatial coordinate, and is polarized in the same direction ${\bf E}=E(z,t){\bf {\hat z}}$. We will refer to this as a 1D electrostatic field geometry, in spite of the time-dependence. Due to the reduced field geometry, it turns out that 8 out of the 16 scalar DHW-variables are zero. Furthermore, due to the linear dependence of some equations, these eight variables can be fully described by four independent quantities and evolution equations \cite{al2021plasma}. How the reduction of the full DHW-theory can be made, without any approximations or assumptions, apart from the direct consequences of the field geometry,  is described in Ref. \cite{al2021plasma}. Here, we will start from the evolution equations describing the electrostatic 1D limit derived in that work. Those equations are:    
\begin{align}
\label{PDE_System}
    D_t\chi_1(z,\bp,t)&= 2\varepsilon_{\bot}(p_{\bot}) \chi_3(z,\bp,t)- \frac{\partial \chi_4}{\partial z} (z,\bp,t)\notag,\\
    D_t\chi_2(z,\bp,t) &= -2p_z\chi_3(z,\bp,t),\\
    D_t\chi_3(z,\bp,t)&= -2\varepsilon_{\bot}(p_{\bot}) \chi_1(z,\bp,t) +2p_z\chi_2(z,\bp,t)\notag,\\
    D_t\chi_4(z,\bp,t)&= -\frac{\partial \chi_1}{\partial z}(z,\bp,t)\notag ,
\end{align}
where $D_t=\partial/\partial t + e\tilde{E}\partial/\partial p_z$ and the non-local electric field operator $\tilde{E}$ is given by
\begin{equation}
\tilde{E}(z)=\int_{-1/2}^{1/2}d\tau \, E(z+i\tau\hbar\partial/\partial p_z). \label{nonlocal}
\end{equation}
Moreover, $\varepsilon_{\perp}=\sqrt{m^2+p_{\perp}^2}$, $\varepsilon=\sqrt{m^2+p_{\perp}^2+p_z^2}$, where $p_{\perp}$ is the perpendicular (to $z$) momentum, and we are using units where $\hbar=c=1$ (we temporarily included $\hbar$ in \cref{nonlocal} to illustrate how quantum effects associated with non-localized particle states enter, but we will leave it out below).   The phase-space variables $\chi_1-\chi_4$ have the following physical interpretations: Firstly, $\chi_1$
gives the (phase space) current density $j_z$, i.e. 
\begin{equation}
    j_z=\frac{e}{(2\pi)^3}\int \chi_{1} \, d^3p.
\end{equation}
Moreover, $\chi_{2}$ gives the mass density $\rho_m$, i.e.
\begin{equation}
    \rho_m =\frac{m}{(2\pi)^3}\int \frac{\chi_{2}}{\varepsilon_{\perp} } \, d^3p.
\end{equation} 
$\chi_{3}$ gives the spin density, i.e. the angular momentum density ${\bf M}$ due to the spin is 
\begin{equation}
    {\bf M} =\frac{1}{(2\pi)^3}\int ({\bf \hat{z}}\times {\bf p}) \frac{\chi_{3}}{2\varepsilon_{\perp} } \, d^3p,
\end{equation} 
and finally $\chi_{4}$ gives the charge density $\rho_c$, i.e. 
\begin{equation}
    \rho_c=\frac{e}{(2\pi)^3}\int \chi_{4} \, d^3p.
\end{equation}
We note that, in addition to the contributions from excitations of the Dirac field, the phase-space variables have the following vacuum contributions, due to the expectation values of the free Dirac field operators (e.g. Ref. \cite{bialynicki1991phase}):  
  \begin{align}
 \chi_{1vac}&=-\frac{2 p_z}{\varepsilon}\notag, \\ \chi_{2vac}&=-\frac{2\varepsilon_{\perp}}{\varepsilon}.
 \end{align}
 To separate real particle contributions as well as field-induced perturbations from the unperturbed vacuum state, we next introduce $\Tilde{\chi}_i(z,\bp,t)$ representing the deviation from the vacuum state, i.e. we let
 \begin{equation}
     \Tilde{\chi}_1(z,\bp,t)=\chi_1(z,\bp,t)- \chi_{1 {\rm vac}}(\bp),
\end{equation}
    and
\begin{equation}
     \Tilde{\chi}_2(z,\bp,t)=\chi_2(z,\bp,t)- \chi_{2 {\rm vac}}(\bp),
    \end{equation}
noting that there is no vacuum contribution to subtract for $\chi_{3,4}$. Next, we assume that the spatial variations occur on a scale much longer than the Compton wavelength, and also much longer than the characteristic de Broglie length of particles. The latter condition allows us to let $\tilde{E}\rightarrow E$ in the evolution equations, such that $D_t=\partial/ \partial t+eE\partial/ \partial p_z$. 
Finally, we make a normalization of variables according to  $t_n= \omega_c t$, $z_n=z \omega_c/c$, $p_{zn}=p_z/mc$, $p_{n\perp}=p_{\perp}/mc$, $E_n=E/E_{cr}$,  $\varepsilon_n=\varepsilon/m$. We note that the DHW-functions are already normalized. With the definitions introduced above, the evolution equations can be written:

\begin{align}
    \label{Eq: Electrostatic Chi1}
    D_t\tilde{\chi}_1 &= 2\varepsilon_\bot\chi_3+2E\frac{\varepsilon_\bot^2}{\varepsilon^3}-\frac{\partial \chi_4}{\partial z},\\
    \label{Eq: Electrostatic Chi2}
    D_t\tilde{\chi}_2 &=-2p_z\chi_3-2p_zE\frac{\varepsilon_\bot}{\varepsilon^3},\\
    \label{Eq: Electrostatic Chi3}D_t\chi_3 &= -2\varepsilon_\bot\tilde{\chi}_1+2p_z\tilde{\chi}_2,\\
    \label{Eq: Electrostatic Chi4}D_t\chi_4 &= -\frac{\partial\tilde{\chi}_1}{\partial z}.
\end{align}
For notational convenience, we have omitted the index $n$ indicating normalized variables from now on. The system is closed with Ampère's law
 which takes the form  
\begin{equation}
\label{Ampers_law}
\frac{\partial E}{\partial t}=-\eta\int \tilde{\chi}_1 \, d^2 p.
\end{equation}
We note that the change $\chi_1\rightarrow\tilde{\chi}_1$ does not affect the momentum-integral, as the unperturbed vacuum background does not contribute, since it is an odd function of momentum. 
Due to the angular symmetry, we have carried out the integration over the azimuthal momentum direction, simply contributing a factor of $2\pi$, and thus the integration measure is $d^2p=p_\perp dp_\perp dp$. Finally, from the normalizations made, the constant $\eta$ gets the value $\eta=\alpha/\pi$ (see e.g. \cite{brodin2023plasma}). \cref{Eq: Electrostatic Chi1,Eq: Electrostatic Chi2,Eq: Electrostatic Chi3,Eq: Electrostatic Chi4} together with (\ref{Ampers_law}), describing the plasma dynamics of the Dirac field in the given field geometry, are the subject of study for the rest of this work.

\subsection{The weak quantum expansion}
In this subsection, we expand the DHW equations in the regime of weak fields (compared to the critical field) and slow time scales (compared to the Compton timescale). In the normalized units, this means $E \ll 1$ and $D_t \ll 1$. Building on this, we use the expansion to derive an evolution equation for the electric field by differentiating Ampère's law with respect to time. During this process, a UV-divergent integral appears, which we handle by renormalizing the charge.

Firstly, we note that since $\chi_3$ is the only inherent quantum DHW-variable (as it is proportional to the spin density), it is useful to express it in terms of the other DHW-variables. By taking the time derivative of Eq.~(\ref{Eq: Electrostatic Chi3}) and using Eqs.~(\ref{Eq: Electrostatic Chi1}) and (\ref{Eq: Electrostatic Chi2}), we arrive at

\begin{equation} \label{Eq: Before trying any expansion}
    \left(D_t^2+4\varepsilon^2\right)\chi_3=-\frac{4\varepsilon_\bot E}{\varepsilon}+2E\tilde{\chi}_2+2\varepsilon_\bot\frac{\partial\chi_4}{\partial z}.
\end{equation}
Now, this can be rewritten as,
\begin{align}
    \chi_3&=D^{-1}\left[\frac{E\tilde{\chi}_2}{2\varepsilon^2}+\frac{\varepsilon_\bot}{2\varepsilon^2}\frac{\partial\chi_4}{\partial z}-\frac{\varepsilon_\bot E}{\varepsilon^3}\right], \label{chi3solved}
\end{align}
where $D\equiv(1/4\varepsilon^2)\left(D_t^2+4\varepsilon^2\right)$ and the inverse operator $D^{-1}$ is given by the Taylor expansion
\begin{align}\label{Eq: Operator D^-1}
   D^{-1}=\left[1-\frac{1}{4\varepsilon^2}D_t^2+\left(\frac{1}{4\varepsilon^2}D_t^2\right)^2-\cdots\right].
\end{align} Since $D_t\ll 1$, the expansion can be truncated to any order, depending on the accuracy needed. 
Next, we make the definitions:
\begin{align}
\label{28}
    \chi_3&=\chi^{(0)}_3+\Delta \chi_3,\\ 
    \chi_3^{(0)}&=\frac{E\tilde{\chi}_2}{2\varepsilon^2}+\frac{\varepsilon_\bot}{2\varepsilon^2}\frac{\partial\chi_4}{\partial z}-\frac{\varepsilon_\bot E}{\varepsilon^3},\\
    \Delta \chi_3&= \left[-\frac{1}{4\varepsilon^2}D_t^2+\left(\frac{1}{4\varepsilon^2}D_t^2\right)^2-\cdots\right]\chi_3^{(0)}.
\end{align}
Moreover, we will use $\Delta \chi_3^{(n)}= (-1)^n\left(\frac{1}{4\varepsilon^2}D_t^2\right)^n\chi_3^{(0)}$, for $n\geq 1$, to denote the $n$:th order term in the expansion of $\chi_3$. 
Keeping terms up to $D_t^4$, i.e. until $n=2$ we end up with
\begin{align}\label{Eq: Chi_3 WCEP expansion}
    \chi_3&=\chi^{(0)}_3+\Delta \chi_3^{(1)}+\Delta\chi_3^{(2)},\\
    \Delta \chi_3^{(1)}&=-\frac{1}{4\varepsilon^2}D_t^2\chi_3^{(0)}, \label{eq27} \\ 
    \Delta\chi_3^{(2)}&=\frac{1}{4\varepsilon^2}D_t^2\left(\frac{1}{4\varepsilon^2}D_t^2\chi_3^{(0)}\right).\label{eq28}
\end{align}
Substituting Eq.~(\ref{Eq: Chi_3 WCEP expansion}) into \cref{Eq: Electrostatic Chi1,Eq: Electrostatic Chi2,Eq: Electrostatic Chi4}, we end up with the following:
\begin{align}
    \label{Eq: WCEP 1}D_t\tilde{\chi}_1&= 2\varepsilon_\bot\left[\frac{E\tilde{\chi}_2}{2\varepsilon^2}-\frac{p_z^2}{2\varepsilon_\bot\varepsilon^2}\frac{\partial\chi_4}{\partial z}+\Delta\chi_3^{(1)}+\Delta\chi_3^{(2)}\right],\\
    \label{Eq: WCEP 2}D_t\tilde{\chi}_2&= -2p_z\left[\frac{E\tilde{\chi}_2}{2\varepsilon^2}+\frac{\varepsilon_\bot}{2\varepsilon^2}\frac{\partial\chi_4}{\partial z}+\Delta\chi_3^{(1)}+\Delta\chi_3^{(2)}\right],\\
    \label{Eq: WCEP 4}
    D_t\chi_4&=-\frac{\partial\tilde{\chi}_1}{\partial z}.
\end{align}
We will refer to the approximation leading up to \cref{Eq: WCEP 1,Eq: WCEP 2,Eq: WCEP 4} as the Weakly Coupled Electron-Positron (WCEP) limit. The name may seem somewhat arbitrary, but it will be justified in the section to follow, when we show that the equations can be rewritten in terms of weakly coupled classical-like distribution functions for electrons and positrons.

\subsection{Charge renormalization} \label{Sec: III C}
In this subsection, we derive an evolution equation for $E$ in the form of a plasma oscillation. In the process, we note the presence of a UV-divergent integral, which we handle via charge renormalization. 

A plasma oscillation equation for $E$ takes the form of a second-order time derivative, involving $\partial_t^2E$. To derive this, we apply the operator $D_t$ to Ampère's law, Eq.~(\ref{Ampers_law}). Combining this with \cref{Eq: WCEP 1} for the right-hand side yields:
\begin{multline}
\frac{\partial^2 E}{\partial t^2}
= -2\eta \int \varepsilon_\bot
\Biggl[
  \frac{E\tilde{\chi}_2}{2\varepsilon^2}
  - \frac{p_z^2}{2\varepsilon_\bot\varepsilon^2}
    \frac{\partial\chi_4}{\partial z}
\\[-4pt]
  + \Delta\chi_3^{(1)}
  + \Delta\chi_3^{(2)}
\Biggr]
\, d^2p.
\label{Renorm-1}
\end{multline}

Now, since $D_t=\partial/\partial t+E\partial/\partial p_z$ we have that $\int D_t\tilde{\chi}_1 \, d^2p=\int \partial_t\tilde{\chi}_1 \, d^2p$ as the momentum derivative becomes zero upon integration. Thus, the left-hand side of Eq.~(\ref{Renorm-1}) is equal to $\int \partial_t\tilde{\chi}_1 \, d^2p$.

We also note that $\Delta\chi_3^{(1)}$ contains a term $\propto(\partial^2 E/\partial t^2) (\varepsilon_\bot/\varepsilon^5)$. When integrating over momentum, this term causes a logarithmic divergence from large momenta, which corresponds to the UV divergences, see e.g. \cite{bialynicki1991phase}. As is well known, this term must be handled by a renormalization. We denote the UV-divergent integral as $I_\mathrm{UV}\equiv \int \frac{\varepsilon_\bot^2}{2\varepsilon^5} \, d^2p$. Moving the UV-divergent term to the left-hand side, \cref{Renorm-1} becomes
\begin{equation}
    \frac{\partial^2 E}{\partial t^2}(1+\eta_B I_\mathrm{UV})=-\eta_B\frac{\partial}{\partial t}\int \tilde{\chi}_{1r} \, d^2p .
\end{equation}
Here we have changed the notation, letting $\eta\rightarrow \eta_B$ as the constant represents the bare value of $\eta$ before the renormalization has been implemented.  Moreover, we have introduced $\chi_{1r}$, which corresponds to the current density with the UV-divergent contribution subtracted.  To identify the renormalized coefficient, we simply write 
\begin{equation} 
  \frac{\partial^2 E}{\partial t^2}=-\eta_R\frac{\partial}{\partial t}\int \tilde{\chi}_{1r} \, d^2p ,\label{renorm-2}
\end{equation}
where  the renormalized coefficient is \begin{equation}
 \eta_R=\frac{\eta_B}{1+\eta_B I_\mathrm{UV}}.
\end{equation}
If we temporarily reinstate the elementary charge, noting that we then have $\eta=\alpha/\pi=e^2/4\pi^2$, we can see that the renormalization of $\eta$ corresponds to a renormalization of the elementary charge, where the bare and normalized charge are related by
\begin{equation}
    e_R^2=\frac{e_B^2}{1+I_\mathrm{UV}e_B^2/4\pi^2}.
\end{equation}
Given \cref{Eq: WCEP 1,Eq: WCEP 2,Eq: WCEP 4,Eq: Chi_3 WCEP expansion,eq27,eq28}, the explicit form of the renormalized \cref{renorm-2}  is given by
\begin{align} \label{Eq: Evolution equation after renormalization}
\frac{\partial^2 E}{\partial t^2} &=     
-\eta_R\int \frac{\varepsilon_\bot}{\varepsilon^2}
\Bigg \{
E\tilde{\chi}_2
- \frac{p_z^2}{\varepsilon_\bot}\frac{\partial\chi_4}{\partial z}+ 2\varepsilon^2\Delta\chi_3^{(2)}
\nonumber\\
& \qquad \qquad \qquad \quad - \frac{1}{2}D_t^2\left(\frac{E\tilde{\chi}_2}{2\varepsilon^2} + \frac{\varepsilon_\bot}{2\varepsilon^2}\frac{\partial \chi_4}{\partial z}\right)
\nonumber\\
&\qquad \qquad  + \frac{3E}{2\varepsilon^5}\left[E^2\left(\frac{5p_z^2}{\varepsilon^2}-1\right)-3p_z\frac{\partial E}{\partial t}\right]
\Bigg \} \, d^2p. 
\end{align}
The evolution of the electric field remains complicated, as it requires the solving of \cref{Eq: WCEP 1,Eq: WCEP 2,Eq: WCEP 4} to proceed. As we will demonstrate in the next section, an improved physical understanding, facilitating the analysis, is helped by a change of variables. 

\section{Quasi-classical distribution functions} \label{Sec: Kinetic theory for electrons and positrons}

In this subsection, we introduce a new set of variables.  The starting point will be our DHW equations, from before we subtracted the vacuum background contributions from $\chi_1$ and $\chi_2$. In combination with the expansion of $\chi_3$ with a vacuum background, we can then derive Vlasov-like equations for these new variables. To the lowest order, we will see that the new variables are governed by Vlasov equations with no source terms, i.e. evolving as classical distribution functions for electrons and positrons \footnote{Here a feature of the DHW-formalism can be noted. While the evolution equation is classical in the $\hbar\rightarrow 0$ limit, the relation between momentum and velocity comes with different signs for particles and antiparticles. Therefore, the convective term of the Vlasov operator, $({p_z}/\varepsilon)\partial/\partial z $ comes with different signs, depending on species.}. Considering higher-order terms, we introduce quantum source terms that couple the two Vlasov-like equations. 

We start with the DHW equations of  (\ref{PDE_System}), and use the third of these equations to directly solve for $\chi_1$:
\begin{equation}
    \chi_1=\frac{p_z\chi_2}{\varepsilon_\bot}-\frac{1}{2\varepsilon_\bot}D_t\chi_3. \label{43}
\end{equation}
Now, we can use the expansion of $\chi_3$ that we found in the previous subsection, i.e. $\chi_3=\chi_3^{(0)}+\Delta\chi_3^{(1)}+\Delta\chi_3^{(2)}$, 
combined with \cref{eq27} and \cref{eq28} such that \cref{43} becomes a solution for $\chi_1$ in terms of $\chi_2$ and $\chi_4$. If we then substitute this solution for $\chi_1$ into \cref{Eq: WCEP 2} and into \cref{Eq: WCEP 4}, we end up with
\begin{align}
D_t \chi_2
  &= -2 p_z \left(
      \frac{E \chi_2 + \varepsilon_\bot \frac{\partial \chi_4}{\partial z}}
           {2 \varepsilon^2}
      + \Delta \chi_3^{(1)}
      + \Delta \chi_3^{(2)}
     \right),
\\
D_t \chi_4
  &= - \frac{1}{\varepsilon_\bot} 
     \frac{\partial}{\partial z} \Bigg[
        p_z \chi_2
\nonumber \\
  &\qquad
        - \frac{1}{2} D_t \left(
            \frac{E \chi_2
                  + \varepsilon_\bot \frac{\partial \chi_4}{\partial z}}
                 {2 \varepsilon^2}
            +  \Delta \chi_3^{(1)}
            +  \Delta \chi_3^{(2)}
          \right)
     \Bigg] .
\end{align}
Now, these will be our starting points for deriving the Vlasov-like equations that we mentioned earlier. To see this, we temporarily drop some terms of the expansion in $\chi_3$, with the purpose of retrieving the classical structure of two Vlasov equations, for electrons and positrons, respectively. In particular, we only keep the $\chi_3^{(0)}$-terms that enter the equations without any $D_t$ acting upon them. To see the classical Vlasov structure, we further introduce the variable $\xi_2\equiv\frac{\varepsilon}{\varepsilon_\bot}\chi_2$. Then the equations take the form
\begin{eqnarray}
D_{t}\left( \frac{\varepsilon _{\bot }}{\varepsilon }\xi _{2}\right)  &=&-\frac{%
p_{z}}{\varepsilon ^{2}}\left( E\frac{\varepsilon _{\bot }}{\varepsilon }\xi
_{2}+\varepsilon _{\bot }\frac{\partial \chi _{4}}{\partial z}\right)  ,\\
D_{t}\chi _{4} &=&-\frac{\partial }{\partial z}\left( \frac{p_{z}\xi _{2}}{%
\varepsilon }\right) .
\end{eqnarray}%
Noting that $D_{t}$ acting on $1/\varepsilon $ in the left-hand side cancels against a term of the right-hand side, we obtain:   
\begin{eqnarray}
D_{t}\xi _{2} &=&-\frac{p_{z}}{\varepsilon }\frac{\partial \chi _{4}}{\partial z%
}, \\
D_{t}\chi _{4} &=&-\frac{\partial }{\partial z}\left( \frac{p_{z}\xi _{2}}{%
\varepsilon }\right) =-\frac{p_{z}}{\varepsilon }\frac{\partial \xi _{2}}{\partial
z}.
\end{eqnarray}%
This gives us two Vlasov equations if we introduce the new variables %
\begin{eqnarray}
f_{+} &=&\xi _{2}+\chi _{4}, \\
f_{-} &=&\xi _{2}-\chi _{4},
\end{eqnarray}%
such that
\begin{eqnarray}
D^+_vf_{+} &=&\left( \frac{\partial}{\partial t}+\frac{p_{z}}{\varepsilon }\frac{\partial }{\partial
z} +E\frac{\partial}{\partial p_z}\right) f_{+}=0, \\
D^-_vf_{-} &=&\left( \frac{\partial}{\partial t}-\frac{p_{z}}{\varepsilon }\frac{\partial }{\partial
z} +E\frac{\partial}{\partial p_z}\right) f_{-}=0,
\end{eqnarray}
which implies a classical evolution of the distribution functions $f_\pm$, with the Vlasov-operators denoted by $D^\pm_v$. Due to the opposite sign in front of the momenta, we interpret the functions as the distribution functions of electrons and positrons \cite{Note1}. We can note that a similar classical structure when letting $\hbar\rightarrow 0$ has been derived in the spinless case, when the Dirac field is replaced by a Klein-Gordon field \cite{best1993phase}. For further comparisons of the DHW-theory and the Klein-Gordon case, see e.g. \cite{al2023applicability}

In the present case, since we defined $e$ as the elementary charge (rather than the electron charge), $D_v^+$ is the Vlasov operator for positrons, and $D_v^-$ is the Vlasov operator for electrons. Next, reinstating the quantum corrections we temporarily omitted, we obtain:
\begin{align}
D^\pm_vf_{\pm} &=-\frac{1}{2\varepsilon_\bot}\Bigg\{4p_{z}\varepsilon \left(\Delta \chi _{3}^{(1)}+\Delta \chi _{3}^{(2)}\right)\notag \\
&\hspace{0.12cm}\pm%
\frac{\partial }{\partial z}\left[ D_{t}\left( 
\frac{E\chi _{2}+\varepsilon _{\bot }\frac{\partial \chi _{4}}{\partial z}}{%
2\varepsilon ^{2}}+\Delta \chi _{3}^{(1)}+\Delta \chi _{3}^{(2)}\right) \right]\Bigg\}.
\label{coupled-vlasov}
\end{align}%
Here, in addition to the Vlasov operators on the left-hand side, the right-hand side now consists of source terms arising from quantum corrections. As a result, $f_\pm$ behave as quasi-classical distribution functions for electrons and positrons, whose evolution includes quantum-corrected coupling terms.

The right-hand side source terms are here still expressed in terms of $\chi _{2}$ and $\chi _{4}$, but we note that they can easily be expressed in terms of the distribution functions using the relations %
\begin{eqnarray}
\chi _{4} &=&\frac{f_{+}-f_{-}}{2}, \\
\chi _{2} &=&\frac{\varepsilon _{\bot }}{\varepsilon }\left( \frac{f_{+}+f_{-}}{2}%
\right), 
\end{eqnarray}%
such that the quasi-classical distribution functions for electrons and positrons couple to each other through the right-hand side source terms. 

Finally, we note that the system is closed using Ampère's law. The current density in terms of the new variables is: 
\begin{align}
\int \chi _{1}d^{3}p &=\int \left( \frac{p_{z}\chi _{2}}{\epsilon _{\bot }}-%
\frac{1}{2\epsilon _{\bot }}D_{t}\chi _{3}\right) \, d^{3}p \nonumber\\
&=\int \left( \frac{p_{z}}{\epsilon }\left( \frac{f_{+}+f_{-}}{2}\right) -%
\frac{1}{2\epsilon _{\bot }}D_{t}\chi _{3}\right) \, d^{3}p .\label{current-dens}
\end{align}%
Since the spin density $\chi_3$ can be expressed in terms of $f_{\pm}$, \cref{coupled-vlasov} together with the current density \cref{current-dens} and Ampere's law constitutes a closed system for $E$ and $f_{\pm}$. We will refer to these equations as the weakly coupled electron-positron (WCEP) equations. A number of features of this system are worth noting:

\begin{enumerate}
\item When deriving the equations for $f_{\pm }$, we did not subtract the
vacuum terms, which are therefore included, due to the contribution from $%
\chi _{2}$. Thus, it may seem as if $f_{\pm }$ have major deviations from
classical distribution functions, as the vacuum contribution is anything but
small. However, if we consider distribution functions with the vacuum value
subtracted, denoted $\tilde{f}_{\pm }$, we note that $D_{v}\tilde{f}_{\pm }=$
$D_{v}f_{\pm }$, which immediately follow from the definitions. Thus, in
order to have distribution functions truly resembling classical
distributions, we may write the equations in terms of $\tilde{f}_{\pm }$
rather than $f_{\pm }$. For this purpose, we note that a simple substitution 
$f_{\pm }\rightarrow \tilde{f}_{\pm }$ works for the left-hand side of \cref{coupled-vlasov}, whereas in the right-hand side we must use 
$\chi _{2}=\varepsilon_{\bot}(\tilde{f}_{+}+%
\tilde{f}_{-})/(2\varepsilon )+\chi _{2vac}$ to express the sources in terms of the classical-like distributions $f_{\pm }$.
\item Besides the polarization current density related to the spin, that is the term proportional to $D_t \chi_3$ in \cref{current-dens},
clearly, the non-classical features come from the coupling terms in the
right-hand side of \cref{coupled-vlasov}. Since all coupling terms come from the spin density $\chi_3$, we can group the terms into three categories, corresponding to the different contributions displayed in the right hand side of \cref{chi3solved}, as we will discuss below.  

\item Firstly, we have terms that do not contain the particle states at
all. This means terms which are not proportional to $\tilde{f}_{\pm }$, but only to $\chi _{2vac}$. We will refer to these terms as "vacuum
polarization terms". Note, however, that the relation to the vacuum
polarization as described in \cref{Sec: II} is not without some ambiguity. To
what degree the terminology "vacuum polarization terms" is appropriate
will be quantified in the next section. 

\item Next we note that for each term proportional to the vacuum mass density, $\chi _{2vac}$, there is also a corresponding term that is instead proportional to the particle mass density,  since $\chi _{2}=\varepsilon_{\bot}(\tilde{f}_{+}+%
\tilde{f}_{-})/(2\varepsilon )+\chi _{2vac}$. We
will label the terms proportional to the particle part of $\chi _{2}$ as
"particle modified vacuum polarization", in case the contribution to $%
\tilde{f}_{+}+\tilde{f}_{-}$ comes from real particles present initially (rather
than from virtual particles contained in $\tilde{f}_{\pm }$, excited from
the vacuum, due to the coupling terms described in point 3). 

\item The third type of terms are those proportional to $\partial \chi _{4}/\partial z$%
, in which case the dependence on the particle distributions $\tilde{f}_{\pm
}$ enters with opposing signs. We will refer to these terms as spin-orbit
corrections, in case the contributions come from particles present
initially. However, we  stress that the relations of such terms to the standard form of
spin-orbit corrections (see e.g. \cite{manfredi2019phase,asenjo2012semi}) are complicated. Nevertheless, terms
of this kind are still proportional to the spin density (since all deviations from classicality come from terms proportional to $\chi_3$, see \cref{coupled-vlasov}, combined with Eqs. (\ref{28})-(\ref{eq28})), are associated with
particle states through $\tilde{f}_{\pm }$, and depend on particle orbits
(position in phase-space), implying that they represent spin-orbit corrections in
a very general sense of the expression. 

\item A somewhat surprising feature of the quantum terms in \cref{coupled-vlasov} is the relation between electron and positron distributions. Typically, when deriving quantum evolution equations without coupling to the quantum vacuum, one starts with a Foldy-Wouthuysen transformation that separates the electron and positron degrees of freedom (see e.g. \cite{asenjo2012semi}). Due to this starting point, the quantum terms in the kinetic evolution equations that are derived (e.g. \cite{manfredi2019phase,asenjo2012semi}) do not couple electrons to positrons. By contrast, for the specific case considered here, we note that the coupling of electrons to positrons is as strong as the quantum self-interaction terms within the same species. This is not a general physical property by any means, but it comes from the fact that certain quantum effects that couple more strongly within the particle species (like e.g. the magnetic dipole force \cite{manfredi2019phase,asenjo2012semi}) vanish in the field geometry of consideration.  
\item When writing the expression for the current density, \cref{current-dens}, we did not address the question of renormalization. However, whether the variables of $f_{\pm}$ or $\chi_{1,2,3,4}$ are used, naturally, the renormalization works the same as described in \cref{Sec: III C}.        
\end{enumerate}
\section{The homogeneous limit} \label{Sec: V}
\subsection{The evolution of the electric field} \label{Sec: V.A}
In this section, we consider the homogeneous case, i.e. when derivatives with respect to $z$ are zero. We begin by imposing this condition on the evolution equation for $E$ and rewrite it using the new variables introduced in the previous section. Next, we derive analytical expressions for $f_\pm$ from the Vlasov-like equations and substitute them back into the evolution equation.

Firstly, we note that in the homogeneous limit $\chi_4=0$, such that $f_+=f_-=f=\xi_2=\frac{\varepsilon}{\varepsilon_\bot}\chi_2$. Next, we let $f=\tilde{f}+f_\mathrm{vac}$ in such a way that
\begin{align}
    \tilde{f} &= \frac{\varepsilon}{\varepsilon_\bot}\tilde{\chi}_2 \label{eqa},\\
    f_\mathrm{vac} &= \frac{\varepsilon}{\varepsilon_\bot}\chi_\mathrm{2vac} = -2 \label{eqb},\\
    \chi_3^{(0)} &= \frac{E\varepsilon_\bot f}{2\varepsilon^3} .\label{eqc}
\end{align}
Moreover, for the homogeneous case, it is convenient to work with canonical momentum variables, as this removes derivatives with respect to momentum in the governing equations, without introducing further complexities. 
Thus, from now on, we introduce the canonical momentum $q_z=p_z+A(t)$, where $A$ is the vector potential. Applying the Weyl gauge (the electrostatic potential is zero, i.e. $E=-d A/dt$), this gives us the simplification $D_t\rightarrow \partial /\partial t$. We note, however, that there is also a cost due to this, since 
$\varepsilon$ gets a time-dependence due to the dependence on the vector potential, i.e. 
$\varepsilon=\varepsilon(t)=\sqrt{1+p_{\perp}^2+(q_z-A(t))^2}$.  

Now, if we introduce canonical momentum, apply \cref{eqa,eqb,eqc} into the evolution equation for $E$, \cref{Eq: Evolution equation after renormalization}, and assume that all derivatives with respect to $z$ are zero, we end up with the following:
\begin{align}
\frac{d^2 E}{dt^2} 
&= -\eta_R \int \frac{\varepsilon_\bot}{\varepsilon^2} 
\Biggl\{
  \frac{E \varepsilon_\bot \tilde{f}}{\varepsilon} 
  - \frac{\varepsilon_\bot}{2} \frac{\partial^2}{\partial t^2} \Big( \frac{E \tilde{f}}{2\varepsilon^3} \Big)+ 2 \varepsilon^2\Delta \chi_3^{(2)} 
\nonumber\\[-2pt]
  & \hspace{-0.25cm}+\frac{3E}{2\varepsilon^5}\left[E^2\left(\frac{5(q_z-A)^2}{\varepsilon^2}-1\right)-3(q_z-A)\frac{dE}{dt}\right]
\Biggr\}
\, d^2q .
\label{eq:long_E}
\end{align}
Moreover, since $f_\mathrm{vac}=-2$, we have that $D_vf_\mathrm{vac}=0$ ($D_v^+=D_v^-=D_v$ in the homogeneous limit also). Taking this into account and imposing the homogeneous limit, we end up with the following equation for $\tilde{f}$,
\begin{equation}
    \frac{\partial \tilde{f}}{\partial t}=-\frac{2(q_z-A)\varepsilon\left(\Delta\chi_3^{(1)}+\Delta\chi_3^{(2)}\right)}{\varepsilon_\bot}. \label{62}
\end{equation}
Now, we assume a slow build-up of the fields over time, starting from an initial state with $E(t=-T_B)=(dE/dt)(t=-T_B)=0$, where the time for build-up $T_B$ is much longer than the inverse Compton frequency. Since the electric field will not grow by itself from such a thermodynamic equilibrium state, we note this will require an external current source in Ampere's law, in addition to the current source from the Dirac field given in \cref{current-dens}. We assume the external current to be turned off after the build-up process, which is assumed to be completed at $t=0$, after which the plasma dynamics involve only the current density from the Dirac field. However, as the right-hand side sources in \cref{62} do not depend explicitly on the current density, only implicitly through the electric field, we can solve for $\tilde{f}$ without specifying the external current density.

Next, using the expansion in weak fields and a slow temporal dependence, we let $\tilde{f}=\tilde{f}^{(0)}+\tilde{f}^{(1)}$, where $\tilde{f}^{(0)}(t=-T_B)$ is the initial distribution of real matter, before the vacuum has been perturbed by electric fields. It is worth noting that a slow temporal dependence requires a limited value of the plasma frequency. Moreover, $\tilde{f}^{(0)}(t=-T_B)$ is proportional to the square of the plasma frequency (after integration over momentum). Thus, as a consequence of the slow temporal dependence, $\tilde{f}^{(0)}(t=-T_B)$ is also considered an expansion parameter. To be concrete, a slow evolution requires the plasma frequency to be much smaller than the Compton frequency. Assuming the above expansion parameters (slow temporal variations, weak electric fields, low plasma frequency) to apply, we can then perturbatively let,
\begin{align}
    \frac{\partial \tilde{f}^{(0)}}{\partial t}&=\frac{(q_z-A)}{2\varepsilon}\frac{\partial^2}{\partial t^2}\left(\frac{E f_\mathrm{vac}}{2\varepsilon^3}\right), \label{63} \\
    \frac{\partial \tilde{f}^{(1)}}{\partial t}&=\frac{(q_z-A)}{2\varepsilon}\frac{\partial^2}{\partial t^2}\Biggl[\frac{E\tilde{f}^{(0)}}{2\varepsilon^3}-\frac{1}{4\varepsilon^2}\frac{\partial^2}{\partial t^2}\left(\frac{E f_\mathrm{vac}}{2\varepsilon^3}\right)\Biggr]. \label{64}
\end{align}
It would be straightforward to go to higher orders in the perturbative expansion, but we choose to truncate at this level of precision, as this will suffice to include vacuum polarization to the leading order. Performing the integration results in the following solutions:
\begin{widetext}
\begin{align}
    \tilde{f}^{(0)} &= \frac{3(q_z-A)^2 E^2}{2\varepsilon^6}+\frac{\varepsilon_\bot^2 E^2}{4\varepsilon^6}-\frac{(q_z-A)} {2\varepsilon^4}\frac{d E}{d t} + \tilde{f}(t=-T_B), \label{fzero}\\ 
    \tilde{f}^{(1)} &= \left(-\frac{1155\varepsilon_\bot^4}{64\varepsilon^{12}}+\frac{63\varepsilon_\bot^2}{2\varepsilon^{10}}-\frac{105}{8\varepsilon^8}\right)E^4 + \left(\frac{105 (q_z-A)}{8\varepsilon^8}-\frac{231(q_z-A)\varepsilon_\bot^2}{16\varepsilon^{10}}  \right)E^2\frac{d E}{d t} + \left(\frac{21\varepsilon_\bot^2}{16\varepsilon^8}-\frac{5}{4\varepsilon^6}\right)\left(\dfrac{d E}{d t}\right)^2\notag\\
    &\quad + \left(\frac{7\varepsilon_\bot^2}{4\varepsilon^8}-\frac{15}{8\varepsilon^6}\right)E\frac{d^2 E}{d t^2} + \frac{(q_z-A)}{8\varepsilon^6}\frac{d^3E}{d t^3} + \left\{\frac{(q_z-A)}{4\varepsilon^4}\frac{d E}{d t}-\frac{E^2}{8\varepsilon^6}\left[6(q_z-A)^2+\varepsilon_\bot^2\right]\right\}\tilde{f}(t=-T_B).
    \label{fone}
\end{align}
Now, before we substitute this back into the evolution equation for $E$, we would like to specify the ordering of the expansion parameters explicitly. For this purpose, let us denote a small parameter by $\delta$. To deduce \cref{fzero,fone} we have assumed that the electric field magnitude and the slowness of temporal variation are of the same order, such that $d/dt  \sim \delta$ and $E\sim\delta$. Furthermore, $\tilde{f}(t=-T_B)\sim\delta^2$ such that $\tilde{f}^{(0)}\sim \delta^2$ and $\tilde{f}^{(1)}\sim \delta^4$. Now, for our evolution equation, we choose to only consider terms up until $\delta^5$. This leaves us with,
\begin{align}
    \frac{d^2 E}{d t^2}&=-\eta_\mathrm{R}\int \frac{\varepsilon_\bot}{\varepsilon^2}\Bigg\{\frac{E\varepsilon_\bot(\tilde{f}^{(0)}+\tilde{f}^{(1)})}{\varepsilon}-\frac{\varepsilon_\bot}{4}\frac{\partial^2}{\partial t^2}\left(\frac{E\tilde{f}^{(0)}}{\varepsilon^3}\right) 
         \nonumber \\
         &\qquad \qquad \qquad \qquad+\frac{3E}{2\varepsilon^5}\left[E^2\left(\frac{5(q_z-A)^2}{\varepsilon^2}-1\right)-3(q_z-A)\frac{d E}{d t}\right]-\frac{\varepsilon_\bot}{8}\frac{\partial^2}{\partial t^2}\left[\frac{1}{\varepsilon^2}\frac{\partial^2}{\partial t^2}\left(\frac{E}{\varepsilon^3}\right)\right]\Bigg\} \, d^2q.
\end{align}
Now, using the expressions for $\tilde{f}^{(0)}$ and $\tilde{f}^{(1)}$, and performing the integrations that does not involve $\tilde{f}(t=-T_B)$, we arrive at,
\begin{align} 
    \frac{d^2E}{d t^2}&=-E\eta_R\int \frac{\varepsilon_\bot^2}{\varepsilon^3}\tilde{f}(t=-T_B) \ d^2q-\frac{2\eta_R}{45}\frac{d^2}{d t^2}\left(E^3\right)+\frac{\eta_R}{15}\frac{d^4E}{d t^4}\notag\\
    &\quad +\eta_R\int \frac{\varepsilon_\bot^2}{4\varepsilon^7}\tilde{f}(t=-T_B)\Bigg\{E^3\left[18\frac{(q_z-A)^2}{\varepsilon^2}+\frac{1}{2}\frac{\varepsilon_\bot^2}{\varepsilon^2}-3\right]+\varepsilon^2\frac{d^2E}{d t^2}-10(q_z-A)E\frac{d E}{d t} \Bigg\} \, d^2q ,
    \label{Ehom-evolution}
\end{align}
\end{widetext}
as the evolution equation for the electric field. Here, we point out that the effect of the renormalization, as described in \cref{Sec: III C}, has been implemented. We note that \cref{Ehom-evolution} has two types of terms. Firstly, the terms that are integrals over the initial distribution of real particles, $\tilde{f}(t=-T_B)$, and secondly, the terms that contribute even in a pure vacuum, which constitute the vacuum polarization. We note that the cubically nonlinear vacuum polarization terms coincide with what we would have obtained from \cref{polarization}. Furthermore, the second vacuum contribution, i.e. the linear term proportional to the fourth-order time derivative, coincides with what we would have obtained from the derivative correction \cref{derivative-corr}, see e.g. Ref. \cite{lundin2010qed} for details. Thus, as intended, we have shown that the DHW formalism recovers known results for the vacuum polarization \cite{marklund2006nonlinear,dittrich2001probing}, while simultaneously including quantum corrections that depend on the matter distribution. 
\subsection{The ultra-relativistic regime}
Next, we analyze \cref{Ehom-evolution} to compare the relative influence of vacuum polarization and matter-dependent quantum corrections. For this purpose, we introduce further simplifications. If the thermal momentum is non-relativistic and the density is low enough to also keep the Fermi-momentum non-relativistic, $\tilde{f}(t=-T_B)$ becomes negligible unless $q_z\ll1$ and $p_{\perp}\ll1$. As a result, we may let $\varepsilon\approx \sqrt{1+A^2}$ and $\varepsilon_\bot \approx 1$ in the integrals over $\tilde{f}(t=-T_B)$.  Accordingly, for Eq.~(\ref{Ehom-evolution}), the only momentum dependence left in the integrals is that of $\tilde{f}(t=-T_B)$. Here, we note that as we initially have only real particles (unperturbed vacuum), the integrals over $\tilde{f}(t=-T_B)$ contribute with a factor proportional to the initial particle density $n_0$, since $\int \tilde{f}(t=-T_B) \ d^2q=n_0$. Moreover, for these densities, corresponding to plasma frequencies much smaller than the Compton frequency, we can deduce that the $d^4E/dt^4$-term of \cref{Ehom-evolution} is negligible \footnote{This can be seen by using the approximation $d^4 E/dt^4 \approx d^2(\omega_{cl}^2E)/d t^2$, and comparing the resulting term with matter dependent quantum corrections.}. 

At this point, 
Eq.~(\ref{Ehom-evolution}) becomes
\begin{align}\label{Esolved}
    \frac{d^2E}{dt^2}\approx&\; \underbrace {-\omega_{cl}^2E}_\text{Classical term}\; + \; \underbrace{- \frac{2\eta_R}{45}\frac{d^2}{dt^2}\left(E^3\right)}_\text{Vacuum polarization contribution}\nonumber\\[6pt]
    &\underbrace{\begin{aligned}
    &+\frac{\omega_{cl}^2}{4(1+A^2)^2}\Bigg[
        E^3\!\left(\frac{18A^2+\tfrac12}{1+A^2}-3\right) \\
    &\qquad\qquad\qquad \qquad
        +\frac{d^2E}{dt^2}
        +10A\,E\,\frac{dE}{dt}
    \Bigg],
\end{aligned}
}_{\text{Matter-dependent correction}}
\end{align}
where we have identified $\omega_{cl}^2=n_0\eta_R/\left(1+A^2\right)^{3/2}$ as the classical contribution \footnote{We can note that prefactor in the classical contribution corresponds to the square of the plasma frequency, but with an extra nonlinear factor $1/(1+A^2)^{3/2}$ due to the relativistic particle motion.}. This is an ODE that, coupled with the Weyl gauge relation $E=-dA/dt$, can be solved for $E(t)$ given initial conditions at $t=0$ for $E, A$ and $dE/dt$ and $n_0$ (related to real particles present before the adiabatic build-up through $\tilde{f}(t=-T_B)$).

Now, we would like to compare the impact of the respective types of quantum corrections (vacuum polarization vs matter-dependent). We denote the classical plasma oscillation period by $T_{cl}$, which we obtain by considering only the classical contributions. Then, for the comparison of the quantum corrections, we examine the deviation from the classical plasma period attained when including either type of correction. Here, we let $T_{vp}$ denote the resulting period when considering the classical contribution and corrections from vacuum polarization. Similarly, when considering the classical contribution and the matter-dependent corrections, we denote the resulting period by $T_{mat}$. 

Moreover, we let $\Delta T_{vp}=T_{vp}-T_{cl}$ and $\Delta T_{mat}=T_{mat}-T_{cl}$ denote the respective deviations from the classical plasma period. Then, the relative impacts on the plasma periods from the quantum corrections, in comparison with the classical period, are given by,
\begin{align}
        R_T^{vp}&=\frac{\Delta T_{vp}}{T_{cl}},\nonumber\\
        R_T^{mat}&=\frac{\Delta T_{mat}}{T_{cl}}.
\end{align}
Also, we compare the impacts of the respective quantum corrections via the ratio $Q_T$ given by,
\begin{equation} \label{Eq: Definition of Q_T}
    Q_T = \Bigg\lvert\frac{R_T^{vp}}{R_T^{mat}}\Bigg\rvert,
\end{equation}
such that $Q_T\geq 1$ happens when the corrections from vacuum polarization dominate over the matter-dependent corrections, and vice versa for $Q_T\leq1$.

For our numerical simulations, we solve the ODE of \cref{Esolved}, in the case of only the classical contribution, classical $+$ vacuum polarization corrections, and classical $+$ matter-dependent corrections. Our initial conditions are given by $E(t=0)=-E_p$ ($E_p$ denotes the peak value of $E$), $A(t=0)=0$, and $(dE/dt)(t=0)=0$ \footnote{Recall that to deduce the solutions given in \cref{fzero} and \cref{fone}, we assumed a slow adiabatic buildup of the electric field. However, once these relations are established, \cref{Esolved} can be solved for arbitrary initial conditions $E(t=0)$, $(dE/dt)(t=0)$, and $A(t=0)$, after the external current due to the adiabatic build-up has been turned off.}. After the electric field has returned to a negative peak, i.e. one full plasma oscillation period, we then document the resulting period. We want to do this for different values of $E_p$ and $\gamma_p$ (peak value of $\gamma$). Now, since $q_z \ll 1$, we have that $|p_z|\approx A$, which means that $\gamma_p\approx A_p$ (peak value of $A$).

\begin{figure}[h!]
    \centering
    \includegraphics[width=\linewidth]{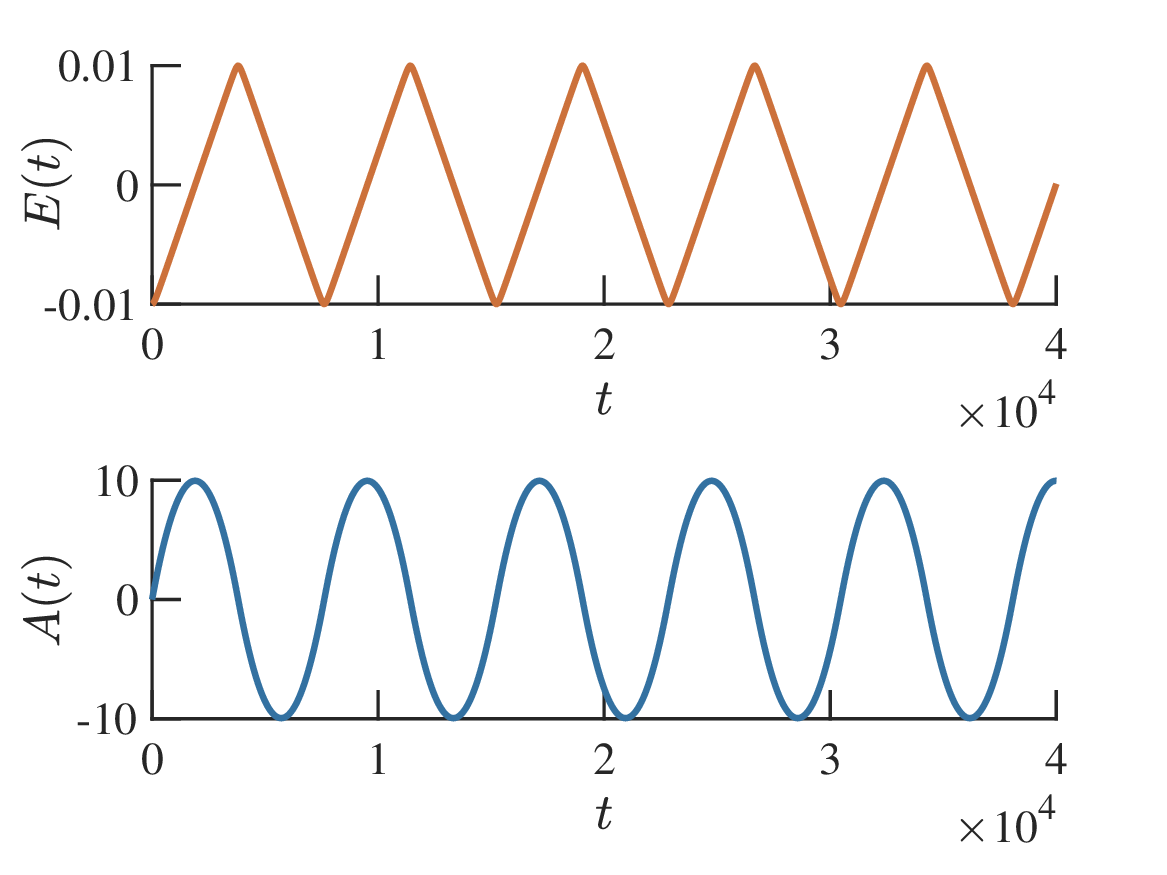}
    \caption{The evolution of $E(t)$ (upper panel) and $A(t)$ (lower panel) based on \cref{Esolved} for the initial conditions $E(t=0)=-0.01$, $A(t=0)=0$, and $(dE/dt)(t=0)=0$. $E_p=0.01$ and $\gamma_p\approx10$. The plasma density, $n_0=\int \tilde{f}(t=-T_B) \, d^2q$, was chosen as $n_0=0.0024$.} 
    \label{fig: EandA for gamma10}
\end{figure}

Now, we choose the values of $E_p$ in the initial conditions. However, for $\gamma_p$ (and thereby also $A_p$), we need a way to control the values we obtain. For this purpose, we note that the electric field profile has an almost constant slope between its negative and positive peaks, which allows us to derive a simple analytical relation between $E_p$, $A_p$, and $n_0$. This property of the ultra-relativistic limit holds regardless of whether we include quantum corrections or not. It stems from the relativistic nonlinearity in the evolution equation of $E$ (due to the $\varepsilon$ in the denominator). Although the assumption of a constant slope becomes increasingly accurate as we approach the ultra-relativistic limit, we already see this behavior for relatively modest values of $\gamma_p$.  In Fig.~(\ref{fig: EandA for gamma10}), we see the electric field $E$ and the vector potential $A$ as functions of time $t$ obtained when solving the ODE with only classical contribution. As can be seen, the triangle-wave-like behavior in $E(t)$ required for our assumptions is clearly visible. By looking at the amplitude of $A$, we see that $\gamma_p \approx 10$. Hence, the constant-slope approximation for $E$ remains accurate even for fairly modest values of $\gamma_p\sim 10$. 

Using this approximation of a constant slope on $E$, we can, by considering the evolution from a negative to a positive peak in $E$, deduce the following relations, $n_0=4E_p/(\eta_RT)$ and $T=8A_p/E_p$, which can be combined to give $A_p=E_p^2/(2\eta_Rn_0)$, 
where $A_p$ fulfills that $A_p\approx\gamma_p$.

\subsection{Numerical Results}


Now, as discussed in the previous subsection, we will solve the ODE of \cref{Esolved} coupled with $E=-dA/dt$ for 
given initial conditions at $t=0$ for $E, A$ and $dE/dt$ and $n_0$, studying the three different cases (classical, classical $+$ vacuum polarization corrections, classical $+$ matter-dependent corrections) over an entire plasma period.  We do this for a range of different values of $E_p$ and $\gamma_p$ (by changing $E_p$ and $n_0$ accordingly). For each case, we document the resulting plasma period. In this subsection, we present the results of these simulations and compare the impact of each quantum correction type. We then determine the regimes where the respective types of correction dominate over the other.



\begin{figure}[h!]
    \centering
    \includegraphics[width=\linewidth]{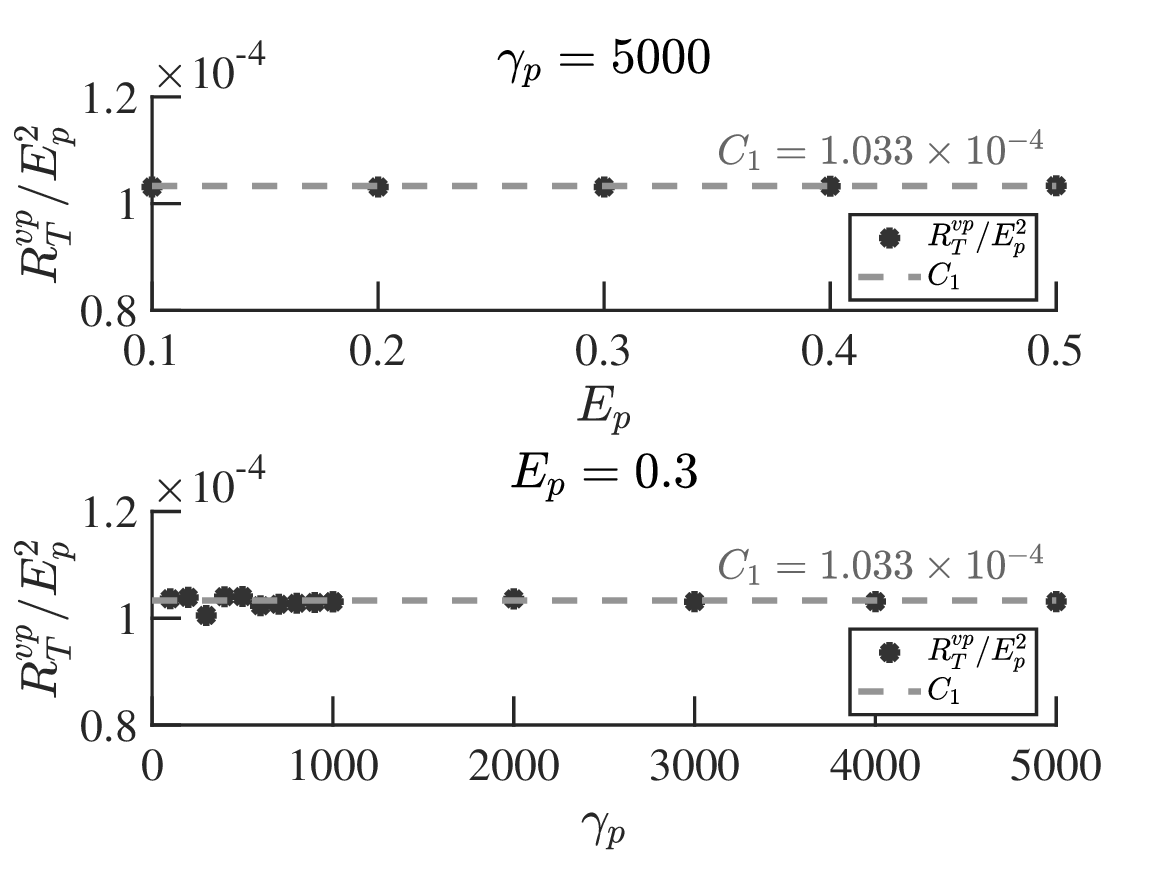}
    \caption{The quantity $R^{vp}_T/E_p^{2}$ (black, dots) compared with the constant value $C_1$ (grey, dashed line). Upper panel: dependence on $E_p$ for fixed $\gamma_p = 5000$. Lower panel: dependence on $\gamma_p$ for fixed $E_p = 0.3$. }
    \label{fig:RT_vp}
\end{figure}

We begin by studying the relative impacts on the plasma periods from the type of quantum corrections, i.e. $R_{T}^{vp}$ and $R_T^{mat}$, separately. Firstly, we consider the deviations in plasma periods that arise when accounting for corrections from vacuum polarization. Fig.~(\ref{fig:RT_vp}) contains two panels of the resulting quantity $R_{T}^{vp}/E_p^2$ from the simulations. In the upper panel of the figure, we keep $\gamma_p=5000$ and vary $E_{p}$. In the lower panel, we instead keep $E_p=0.3$ and vary $\gamma_p$. We include a dashed line equal to $C_1=1.033\times 10^{-4}$ in both panels. In both cases, we see that the quantity $R_{T}^{vp}/E_p^2$ remains approximately constant and is equal to $C_1$. This tells us that $R_{T}^{vp}\approx C_1E_p^2$. 

\begin{figure}[htb!]
    \centering
    \includegraphics[width=\linewidth]{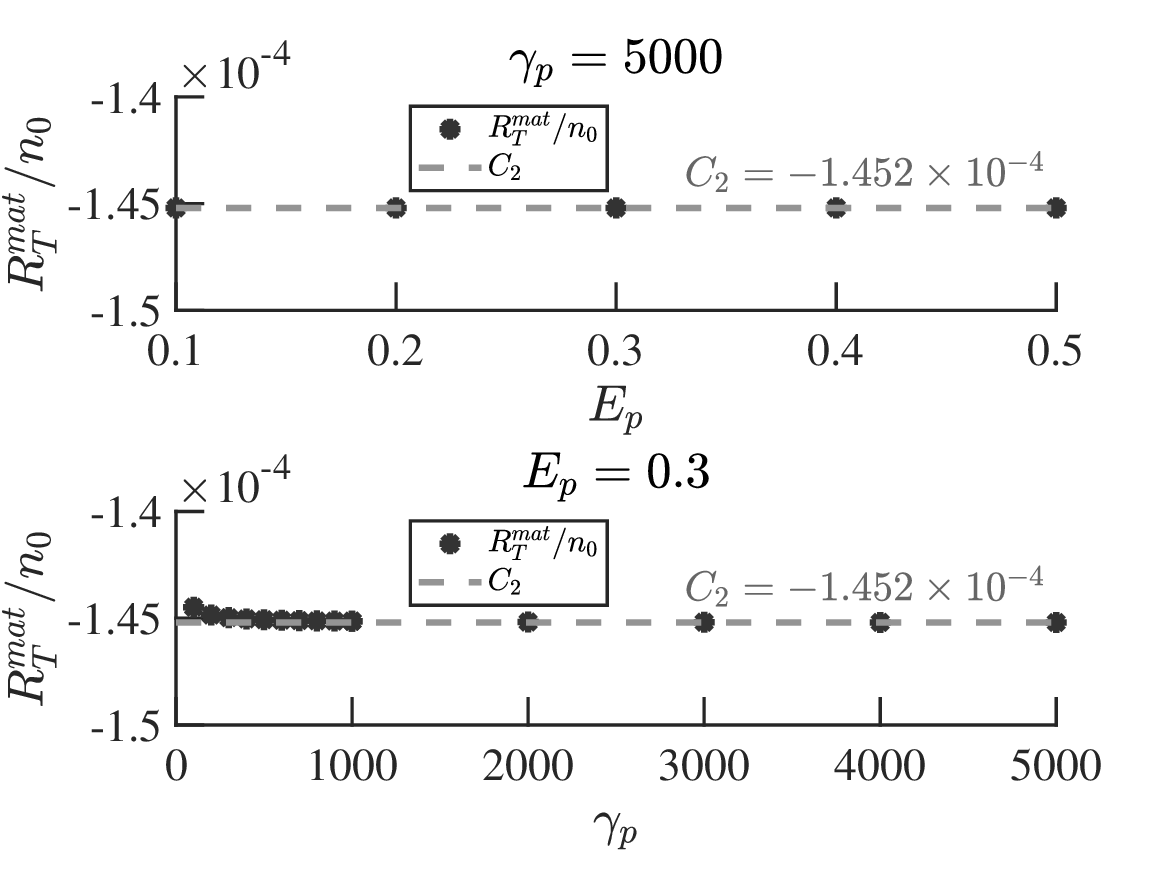}
    \caption{The quantity $R^{mat}_T/n_0$ (black, dots) compared with the constant value $C_2$ (grey, dashed line). Upper panel: dependence on $E_p$ for fixed $\gamma_p = 5000$. Lower panel: dependence on $\gamma_p$ for fixed $E_p = 0.3$.}
    \label{fig:RT_mat}
\end{figure}

In a similar fashion, we then instead consider quantum corrections that stem from the matter-dependent terms. Here, Fig.~(\ref{fig:RT_mat})  displays two similar panels, but this time for the quantity $R_{T}^{mat}/n_0$ along with the constant dashed line equal to $C_2=-1.452\times 10^{-4}$. From the figure, we see that the quantity stays approximately constant and follows the dashed line, thus telling us that $R_{T}^{mat}\approx C_2n_0$.



Now, using the results for $R_T^{vp}$ and  $R_T^{mat}$, we deduce that
\begin{equation}
    Q_T\approx\Bigg\lvert \frac{C_1}{C_2} \frac{E_{p}^2}{n_0} \Bigg\rvert,
\end{equation}
where we use the relation $\gamma_p\approx A_p\approx\frac{1}{2\eta_\mathrm{R}}\frac{E_p^2}{n_0}$, to arrive at
\begin{equation}
Q_T\approx\Bigg\lvert2 \frac{C_1}{C_2} \eta_\mathrm{R}\gamma_p\Bigg\rvert,
\end{equation}
such that we have $Q_T\geq 1$ for $\gamma_p \geq (1.452\times 137\pi)/(2\times1.033)\approx 302 \sim 300$. This means that for $\gamma_p \gtrsim 300$ vacuum polarization becomes the dominant quantum correction. While this result has been derived for a simplified case, to a certain degree, we expect the scaling with the relativistic factor to be a general feature. In particular, also for more general cases, we expect an increasing relativistic factor to contribute to a larger effective mass of real particles, effectively reducing the real particle currents, including the quantum corrections. 






\section{Summary and conclusion} \label{Sec: VI}



In the present paper, we have studied the combined effects of vacuum polarization and quantum corrected matter currents, in the quantum kinetic framework of the DHW formalism, for the case of 1D electrostatic fields. For field strengths smaller than the critical field, and time scales longer than the inverse Compton frequency, it turns out that the DHW equations can be separated into classical-like distribution functions for electrons and positrons, weakly coupled through the quantum vacuum, and general forms of spin-orbit interaction. The resulting  \cref{coupled-vlasov} coupled through the current density given by \cref{current-dens} have here been referred to as the WCEP equations. 

Analyzing strongly relativistic plasma oscillations, the wave motion is affected by vacuum polarization, with the corresponding expressions agreeing with results deduced from the Euler-Heisenberg Lagrangian, including derivative corrections. The vacuum corrections are complemented by quantum corrections affecting the particle currents. Comparing the change in the plasma oscillation period, arising from the vacuum polarization and the matter quantum corrections respectively, we deduce that the vacuum polarization dominates in the ultra-relativistic regime, specifically when $\gamma\gtrsim300$, and the matter-dependent quantum corrections dominate elsewhere. For the specific case considered in the present paper, and for a given electric field strength, it is the plasma density that determines the wave frequency and thereby the magnitude of the relativistic factor. Thus, as one might intuitively expect, letting a classical model for the current density be complemented with quantum corrections due to vacuum polarization will be a feasible approximation for a sufficiently low plasma density. 

We emphasize that the vacuum polarization comes directly from the coupling terms in the kinetic evolution equations for the electrons and positrons. An important feature is that the lowest-order vacuum corrections in the evolution equation cancel when integrating over momentum to form the vacuum polarization current density. When solving for the particle distribution functions, the phase-space densities get contributions from the quantum vacuum of the order $E^2$ (cf. \cref{fzero}). However, terms of that order (as well as all terms of larger order than $E^2d E/d t)$, cancel when computing the vacuum polarization current density. In the present calculations, we have not seen a distinct dynamical influence of the largest coupling terms. Roughly, one can interpret this as virtual electrons and positrons being created in equal measure, not producing charge or current densities; hence, higher-order terms are needed to exert a dynamical influence on the electric field evolution. 

However, the presence of matter affects the vacuum contribution. Specifically, the vacuum modification of the phase-space density differs depending on whether there is a non-zero phase-space density of real particles at a given position in phase space. This is seen in \cref{Ehom-evolution}, where terms proportional to matter give quantum corrections of order $E^3$, while the vacuum polarization terms are of order $d^2 (E^3)/d t^2$.  
We note that when studying phenomena like wave-particle interaction (for wave-particle interaction in a DHW context, see e.g. \cite{al2022linear}), typically, the phase-space density at a specific position in phase-space is crucial. Thus, the larger vacuum modifications of the phase space density ($\propto E^2$), are not necessarily small compared to the phase-space density of real particles. As a result, such virtual particle generation may be dynamically important, although the effect does not correspond to real particle generation. 

The approach of the present paper, studying the DHW equations to derive evolution equations for weakly coupled electrons and positrons, can be generalized to the case of arbitrary field geometries. Phenomena involving wave-particle interaction from cyclotron resonances would then be of particular interest, as the magnitude of such phenomena depends on the phase-space density at given positions in phase space, where the interplay between the quantum vacuum and matter quantum effects can be decisive. However, the case of a more general field geometry is a project for future research.  

\section{acknowledgment}
The author Haidar Al-Naseri acknowledges support from the Knut and Alice Wallenberg Foundation under project number KAW 2022.0361.

\bibliography{References}   

@article{Stratonovich1956Dokl,
  author       = {Stratonovich, R. L.},
  title        = {A gauge-invariant analog of the Wigner distribution},
  journal      = {Doklady Akademii Nauk SSSR},
  volume       = {109},
  pages        = {72--75},
  year         = {1956},
  note         = {In Russian}
}

@article{lai2003transfer,
  title={Transfer of polarized radiation in strongly magnetized plasmas and thermal emission from magnetars: effect of vacuum polarization},
  author={Lai, Dong and Ho, Wynn CG},
  journal={The Astrophysical Journal},
  volume={588},
  number={2},
  pages={962},
  year={2003},
  publisher={IOP Publishing}
}

@article{eriksson2004possibility,
  title={Possibility to measure elastic photon-photon scattering in vacuum},
  author={Eriksson, Daniel and Brodin, Gert and Marklund, Mattias and Stenflo, Lennart},
  journal={Physical Review A—Atomic, Molecular, and Optical Physics},
  volume={70},
  number={1},
  pages={013808},
  year={2004},
  publisher={APS}
}

@article{vladimirov2011description,
  title={On description of a collisionless quantum plasma},
  author={Vladimirov, Sergei V and Tyshetskiy, Yu O},
  journal={Physics-Uspekhi},
  volume={54},
  number={12},
  pages={1243},
  year={2011},
  publisher={IOP Publishing}
}

@article{brodin2022quantum,
  title={Quantum kinetic theory of plasmas},
  author={Brodin, Gert and Zamanian, Jens},
  journal={Reviews of Modern Plasma Physics},
  volume={6},
  number={1},
  pages={4},
  year={2022},
  publisher={Springer}
}

@article{melrose2020quantum,
  title={Quantum kinetic theory for unmagnetized and magnetized plasmas: A tutorial review of quantum plasma theory},
  author={Melrose, DB},
  journal={Reviews of Modern Plasma Physics},
  volume={4},
  number={1},
  pages={8},
  year={2020},
  publisher={Springer}
}

@article{manfredi2021fluid,
  title={Fluid descriptions of quantum plasmas},
  author={Manfredi, Giovanni and Hervieux, Paul-Antoine and Hurst, J{\'e}r{\^o}me},
  journal={Reviews of Modern Plasma Physics},
  volume={5},
  number={1},
  pages={7},
  year={2021},
  publisher={Springer}
}

@article{gonoskov2022charged,
  title={Charged particle motion and radiation in strong electromagnetic fields},
  author={Gonoskov, A and Blackburn, TG and Marklund, M and Bulanov, SS},
  journal={Reviews of Modern Physics},
  volume={94},
  number={4},
  pages={045001},
  year={2022},
  publisher={APS}
}

@article{harding2006physics,
  title={Physics of strongly magnetized neutron stars},
  author={Harding, Alice K and Lai, Dong},
  journal={Reports on Progress in Physics},
  volume={69},
  number={9},
  pages={2631},
  year={2006},
  publisher={IOP Publishing}
}

@article{rozanov1998self,
  title={Self-action of intense electromagnetic radiation in an electron-positron vacuum},
  author={Rozanov, NN},
  journal={Journal of Experimental and Theoretical Physics},
  volume={86},
  number={2},
  pages={284--288},
  year={1998},
  publisher={Springer}
}

@article{lundin2007short,
  title={Short wavelength electromagnetic propagation in magnetized quantum plasmas},
  author={Lundin, Joakim and Zamanian, Jens and Marklund, Mattias and Brodin, Gert},
  journal={Physics of plasmas},
  volume={14},
  number={6},
  year={2007},
  publisher={AIP Publishing}
}

@article{mamaev1981effective,
  title={Effective action for a non-stationary electromagnetic field},
  author={Mamaev, SG and Mostepanenko, VM and Eides, Michael I},
  journal={Sov. J. Nucl. Phys},
  volume={33},
  number={4},
  pages={569--572},
  year={1981}
}

@article{ilderton2012lightfront,
  title={Lightfront DHW functions and strong field QED},
  author={Ilderton, Anton},
  journal={Few-body systems},
  volume={52},
  number={3},
  pages={431--436},
  year={2012},
  publisher={Springer}
}

@article{asenjo2012semi,
  title={Semi-relativistic effects in spin-1/2 quantum plasmas},
  author={Asenjo, Felipe A and Zamanian, Jens and Marklund, Mattias and Brodin, Gert and Johansson, Petter},
  journal={New Journal of Physics},
  volume={14},
  number={7},
  pages={073042},
  year={2012},
  publisher={IOP Publishing}
}

@article{hebenstreit2010schwinger,
  title={Schwinger pair production in space-and time-dependent electric fields: Relating the Wigner formalism to quantum kinetic theory},
  author={Hebenstreit, Florian and Alkofer, Reinhard and Gies, Holger},
  journal={Physical Review D—Particles, Fields, Gravitation, and Cosmology},
  volume={82},
  number={10},
  pages={105026},
  year={2010},
  publisher={APS}
}

@book{dittrich2001probing,
  title={Probing the quantum vacuum: pertubative effective action approach in quantum electrodynamics and its application},
  author={Dittrich, Walter and Gies, Holger},
  year={2001},
  publisher={Springer}
}

@article{manfredi2019phase,
  title={Phase-space modeling of solid-state plasmas: A journey from classical to quantum},
  author={Manfredi, Giovanni and Hervieux, Paul-Antoine and Hurst, J{\'e}r{\^o}me},
  journal={Reviews of Modern Plasma Physics},
  volume={3},
  number={1},
  pages={13},
  year={2019},
  publisher={Springer}
}

@article{brodin2023plasma,
  title={Plasma dynamics at the Schwinger limit and beyond},
  author={Brodin, Gert and Al-Naseri, Haidar and Zamanian, Jens and Torgrimsson, Greger and Eliasson, Bengt},
  journal={Physical Review E},
  volume={107},
  number={3},
  pages={035204},
  year={2023},
  publisher={APS}
}

@article{andreev2015separated,
  title={Separated spin-up and spin-down quantum hydrodynamics of degenerated electrons: Spin-electron acoustic wave appearance},
  author={Andreev, Pavel A},
  journal={Physical Review E},
  volume={91},
  number={3},
  pages={033111},
  year={2015},
  publisher={APS}
}

@article{al2021plasma,
  title={Plasma dynamics and vacuum pair creation using the Dirac-Heisenberg-Wigner formalism},
  author={Al-Naseri, Haidar and Zamanian, Jens and Brodin, Gert},
  journal={Physical Review E},
  volume={104},
  number={1},
  pages={015207},
  year={2021},
  publisher={APS}
}

@article{hussain2014weakly,
  title={Weakly relativistic quantum kinetic theory for electrostatic wave modes in magnetized plasmas},
  author={Hussain, Azhar and Stefan, Martin and Brodin, Gert},
  journal={Physics of Plasmas},
  volume={21},
  number={3},
  year={2014},
  publisher={AIP Publishing}
}

@article{li2019ultrarelativistic,
  title={Ultrarelativistic electron-beam polarization in single-shot interaction with an ultraintense laser pulse},
  author={Li, Yan-Fei and Shaisultanov, Rashid and Hatsagortsyan, Karen Z and Wan, Feng and Keitel, Christoph H and Li, Jian-Xing},
  journal={Physical review letters},
  volume={122},
  number={15},
  pages={154801},
  year={2019},
  publisher={APS}
}

@article{del2018electron,
  title={Electron spin polarization in realistic trajectories around the magnetic node of two counter-propagating, circularly polarized, ultra-intense lasers},
  author={Del Sorbo, Dario and Seipt, Daniel and Thomas, Alexander GR and Ridgers, CP},
  journal={Plasma Physics and Controlled Fusion},
  volume={60},
  number={6},
  pages={064003},
  year={2018},
  publisher={IOP Publishing}
}

@article{del2017spin,
  title={Spin polarization of electrons by ultraintense lasers},
  author={Del Sorbo, Dario and Seipt, Daniel and Blackburn, Tom G and Thomas, Alexander GR and Murphy, Christopher D and Kirk, John G and Ridgers, CP},
  journal={Physical Review A},
  volume={96},
  number={4},
  pages={043407},
  year={2017},
  publisher={APS}
}

@article{di2012extremely,
  title={Extremely high-intensity laser interactions with fundamental quantum systems},
  author={Di Piazza, A and M{\"u}ller, C and Hatsagortsyan, KZ},
  journal={Reviews of Modern Physics},
  volume={84},
  number={3},
  pages={1177--1228},
  year={2012},
  publisher={APS}
}

@article{fedotov2023advances,
  title={Advances in QED with intense background fields},
  author={Fedotov, A and Ilderton, A and Karbstein, F and Seipt, D and Taya, H},
  journal={Physics Reports},
  volume={1010},
  pages={1--138},
  year={2023},
  publisher={Elsevier}
}

@article{chen2013nonlinear,
  title={Nonlinear theory of intense laser-plasma interactions modified by vacuum polarization effects},
  author={Chen, Wenbo and Bu, Zhigang and Li, Hehe and Luo, Yuee and Ji, Peiyong},
  journal={Physics of Plasmas},
  volume={20},
  number={7},
  year={2013},
  publisher={AIP Publishing}
}

@article{medvedev2023plasma,
  title={Plasma modes in QED super-strong magnetic fields of magnetars and laser plasmas},
  author={Medvedev, Mikhail V},
  journal={Physics of Plasmas},
  volume={30},
  number={9},
  year={2023},
  publisher={AIP Publishing}
}

@article{marklund2006nonlinear,
  title={Nonlinear collective effects in photon-photon and photon-plasma interactions},
  author={Marklund, Mattias and Shukla, Padma K},
  journal={Reviews of modern physics},
  volume={78},
  number={2},
  pages={591--640},
  year={2006},
  publisher={APS}
}

@article{heyl1999nonlinear,
  title={Nonlinear QED effects in strong-field magnetohydrodynamics},
  author={Heyl, Jeremy S and Hernquist, Lars},
  journal={Physical Review D},
  volume={59},
  number={4},
  pages={045005},
  year={1999},
  publisher={APS}
}

@book{melrose2013qpd2,
  author    = {D. B. Melrose},
  title     = {Quantum Plasmadynamics: Magnetized Plasmas},
  year      = {2013},
  publisher = {Springer},
  address   = {New York},
  series    = {Lecture Notes in Physics},
  volume    = {870},
  doi       = {10.1007/978-1-4614-6061-3},
  isbn      = {978-1-4614-6060-6},
}

@article{huang2015quantum,
  title={Quantum-electrodynamical birefringence vanishing in a thermal relativistic pair plasma},
  author={Huang, YS},
  journal={Scientific Reports},
  volume={5},
  number={1},
  pages={15866},
  year={2015},
  publisher={Nature Publishing Group UK London}
}

@article{qu2021signature,
  title={Signature of collective plasma effects in beam-driven QED cascades},
  author={Qu, Kenan and Meuren, Sebastian and Fisch, Nathaniel J},
  journal={Physical review letters},
  volume={127},
  number={9},
  pages={095001},
  year={2021},
  publisher={APS}
}

@article{sheng2019wigner,
  title={Wigner function and pair production in parallel electric and magnetic fields},
  author={Sheng, Xin-li and Fang, Ren-hong and Wang, Qun and Rischke, Dirk H},
  journal={Physical Review D},
  volume={99},
  number={5},
  pages={056004},
  year={2019},
  publisher={APS}
}

@phdthesis{lundin2010qed,
  title={QED and collective effects in vacuum and plasmas},
  author={Lundin, Joakim},
  year={2010},
  school={Ume{\aa} universitet. Institutionen f{\"o}r fysik}
}

@article{bialynicki1991phase,
  title={Phase-space structure of the Dirac vacuum},
  author={Bialynicki-Birula, Iwo and Gornicki, Pawel and Rafelski, Johann},
  journal={Physical Review D},
  volume={44},
  number={6},
  pages={1825},
  year={1991},
  publisher={APS}
}

@article{al2022linear,
  title={Linear pair-creation damping of high-frequency plasma oscillation},
  author={Al-Naseri, Haidar and Brodin, Gert},
  journal={Physics of Plasmas},
  volume={29},
  number={4},
  year={2022},
  publisher={AIP Publishing}
}

@article{best1993phase,
  title={The phase-space structure of the Klein-Gordon field},
  author={Best, Christoph and Gornicki, Pawel and Greiner, Walter},
  journal={Annals of Physics},
  volume={225},
  number={2},
  pages={169--190},
  year={1993},
  publisher={Elsevier}
}

@article{al2023applicability,
  title={Applicability of the Klein-Gordon equation for pair production in vacuum and plasma},
  author={Al-Naseri, Haidar and Brodin, Gert},
  journal={Physical Review E},
  volume={108},
  number={5},
  pages={055205},
  year={2023},
  publisher={APS}
}

\end{document}